\documentclass[useAMS,usenatbib]{mn2e}
\usepackage[letterpaper, totalwidth=480pt, totalheight=680pt]{geometry}
\usepackage{longtable}
\usepackage{graphicx}

\title[The CoRoT star 105288363]{The CoRoT star 105288363: strong cycle to cycle changes of the Blazhko modulation}
\author[E. Guggenberger, K. Kolenberg,  E. Chapellier, E. Poretti et al.]{E. Guggenberger$^{1}$\thanks{E-mail:
elisabeth.guggenberger@univie.ac.at}, K. Kolenberg$^{1,2}$, E. Chapellier$^{3}$, E. Poretti$^{4}$, R. Szab\'o$^{5}$ , 
\and J.M. Benk\H{o}$^{5}$, M. Papar\'o$^{5}$\\
$^{1}$Institut f\"ur Astronomie, Universit\"at Wien, T\"urkenschanzstrasse 17, A-1180 Vienna, Austria\\
$^{2}$Harvard-Smithsonian Center for Astrophysics, 60, Garden street, Cambridge MA 02138, USA\\ 
$^{3}$Observatoire de la C$\hat{o}$te d'Azur, Universite Nice Sophia-Antipolis, UMR 6525, Parc Valrose, 06108 Nice Cedex 02, France\\
$^{4}$INAF- Osservatorio Astronomico di Brera, Via E. Bianchi 46, 23807 Merate (LC), Italy\\
$^{5}$Konkoly Oberservatory of the Hungarian Academy of Sciences, PO Box 67, H-1525 Budapest, Hungary}
\begin{document}

\date{Accepted 0000, Received 0000; in original form 0000 }

\pagerange{\pageref{firstpage}--\pageref{lastpage}} \pubyear{0000}

\maketitle

\label{firstpage}

\begin{abstract}
We present the analysis of the CoRoT star 105288363, a new Blazhko RR Lyrae star of type RRab ($f_0=1.7623$ d$^{-1}$), observed with the CoRoT space craft during the second long run in direction of the galactic center (LRc02, time base 145~d). The CoRoT data are characterized by an excellent time sampling and a low noise amplitude of 0.07 mmag in the 2-12 d$^{-1}$ range and allow us to study not only the fine details of the variability of the star but also long-term changes in the pulsation behaviour and the stability of the Blazhko cycle. We use, among other methods, standard Fourier analysis techniques and O-C diagrams to investigate the pulsational behavior of the Blazhko star 105288363. In addition to the frequency pattern expected for a Blazhko RR Lyrae star, we find an independent mode ($f_1=2.984$ d$^{-1}$) showing a $f_0/f_1$ ratio of 0.59 which is similar to that observed in other Blazhko RR Lyrae stars. The bump and hump phenomena are also analysed, with their variations over the Blazhko cycle. We carefully investigated the strong cycle-to-cycle changes in the Blazhko modulation ($P_B=35.6$ d), which seem to happen independently and partly diametrically in the amplitude and the phase modulation. Furthermore, the phasing between the two types of modulation is found to change during the course of the observations.
\end{abstract}

\begin{keywords}
techniques: photometric, stars: variables: RR Lyrae: individual: CoRoT-ID 105288363, 
\end{keywords}

\section{Introduction}

RR Lyrae stars are low-mass stars on the horizontal branch that burn Helium in their core. They have contributed to the progress in many fields of modern astronomy. With their large amplitudes of up to 1.5 magnitudes they have been known since the end of the XIX century and offer the opportunity to study stellar evolution in real-time \citep{LeBo07}. They obey a period-luminosity-color relation which makes them important standard candles. Also, their evolutionary status makes them useful for the study of galactic evolution. RR Lyrae stars can pulsate in the radial fundamental mode (RRab stars), in the first radial overtone (RRc stars) or in both modes simultaneously (RRd stars). Some RR Lyrae stars are candidates for possible higher radial overtone pulsation.\\

One of the remaining unsolved questions in those seemingly well-studied stars is the century-old Blazhko enigma, a modulation of the amplitude and/or phase of the main pulsation period on time scales of typically weeks to months. With the availability of new and precise data, the estimated incidence rate of this phenomenon has recently increased dramatically, making it even more important to understand the effect. While until some years ago, a value of 20-30 per cent was assumed for RRab stars, \citet{kol10a} find at least 40 per cent of the stars observed by the Kepler satellite to be modulated, and \citet{jur09} give a lower limit of 50 per cent of modulated stars in their sample and even raise the suspicion that the modulation might be a universal property of RRab stars.\\

Several promising models have been proposed to explain the Blazhko phenomenon, among those are the magnetic oblique pulsator-rotator model \citep[e.g.,][]{shi} that works similarly to the mechanism that causes the typical patterns of roAp variation, as well as models involving resonances between the main pulsation and either higher-order radial modes or non-radial modes \citep[e.g.,][and references therein]{dzi}. None of these models, however, succeeds in explaining the full spectrum of observed features of Blazhko stars. The repeated non-detection of the required strong dipole magnetic fields in RR Lyrae stars in recent studies \citep{cha04, schne, kol09, gug09} poses a significant problem for the magnetic models, and the changes that have been reported to occur in the Blazhko phenomenon of several stars, especially variations of its period are a big challenge for any model that links the Blazhko effect directly to rotation and therefore predicts an almost clock-work like behaviour.\\

A scenario that explicitly favors irregular or stochastic Blazhko cycles was published by \citet{sto}. It connects the amplitude and phase modulations of Blazhko stars to a cyclic weakening and strengthening of the turbulent convection inside the helium and hydrogen ionisation zones due to the presence of local transient magnetic fields.\\

Most recently, the detection of period doubling in several RR Lyrae stars in the high-precision data delivered by the Kepler mission \citep{kol10b, szabo10} triggered the exploration of radial resonances in Blazhko RR Lyrae stars. While hydrodynamical simulations successfully reproduced period doubling \citep{szabo10, kollath11}, and found a 9:2 resonance between the 9th overtone which appears as a surface mode and the fundamental pulsation to be responsible for the phenomenon, they were not able to yield modulated light curves. It was an alternative approach by \citet{buchler11}, using the amplitude equation formalism, that succeeded in producing amplitude modulation of both regular and irregular type. In their simulations, chaos occurs because of the presence of a strange attractor in the dynamics, causing irregular behaviour.\\

Changes of both the amplitude and/or the length of the Blazhko cycles, irregularities in the Blazhko phenomenon, sometimes also accompanied by a change in the main pulsation period have been reported for a sample of RR~Lyrae stars. A detailed look at the known changes in Blazhko cycles of different stars will be taken in Section \ref{changes}. \\
One has to note, however, that previous studies of changing Blazhko phenomena usually relied on ground-based data only, suffering from small or sometimes large gaps due to daylight and/or weather, and therefore also often from incomplete coverage of the pulsation and especially the Blazhko period. Reports on long-term variations of the Blazhko effect in specific stars also often compare publications of different authors using different methods with varying precisions. It was therefore often impossible to say when the changes took place and whether they happened abruptly or continuously. In this study, we have, for the first time, investigated strong cycle-to-cycle changes of the Blazhko effect in a continuous and homogeneous data set that covers 255 pulsation and more than 4 full Blazhko cycles.\\

Section \ref{data} will present the data used for analysis, as well as data processing methods. Section \ref{analysis} explains the analysis methods and presents the results, while their implications are discussed in Section~\ref{disc}.
At the final stages of preparing our manuscript, we were informed that another group had analyzed the same data set and published their results in \citet{cha11}. The main differences between the results of the two manuscripts are briefly summarized in appendix \ref{app}.\\

\section{The Target}
The star CoRoT 105288363 $(\alpha=18^{h} 39^{m} 30.8^{s},  \delta=+7\degr 26' 53.9'', J2000$) is a 15.3 magnitude star (V) in the constellation Ophiuchus. Before the CoRoT mission, it was not known to be variable. CoRoT 105288363 was classified as an RR~Lyrae variable with the 'CoRoT Variability Classifier', using two different automated supervised classification methods \citep{deb} and was then confirmed by human inspection. The ExoDat database \citep{del}, which provides important information on the parameters of the stars observed in the CoRoT exoplanet field, lists a contamination of the CoRoT 105288363 data of 0.177, meaning that 17.7 per cent of the measured flux might originate from background stars. The nearest background stars also included in the photometric mask of CoRoT 105288363 are also listed in ExoDat and are all 2 or more magnitudes fainter than the target star.\\

\section{Observations and data treatment}
\label{data}
The data used in this study were obtained in the framework of the Additional Programmes of the exoplanetary field  during the second long run in direction galactic center (LRc02). Observations were carried out during a time span of 145 d from April 15, 2008 to September 7, 2008, or CoRoT JD 3027.47 to 3172.46. Note that the zero point of CoRoT JD corresponds to HJD 2451545.0, which is January 1, 2000 at 12:00:00 UT. The nominal exposure time in the exoplanet field is 512 s, as 16 measurements with a sampling of 32 s are averaged on board \citep{deb}. For this study we used the calibrated light curves of N2 level.\\

\begin{figure}
\includegraphics[width=90mm, bb= 10 0 490 570]{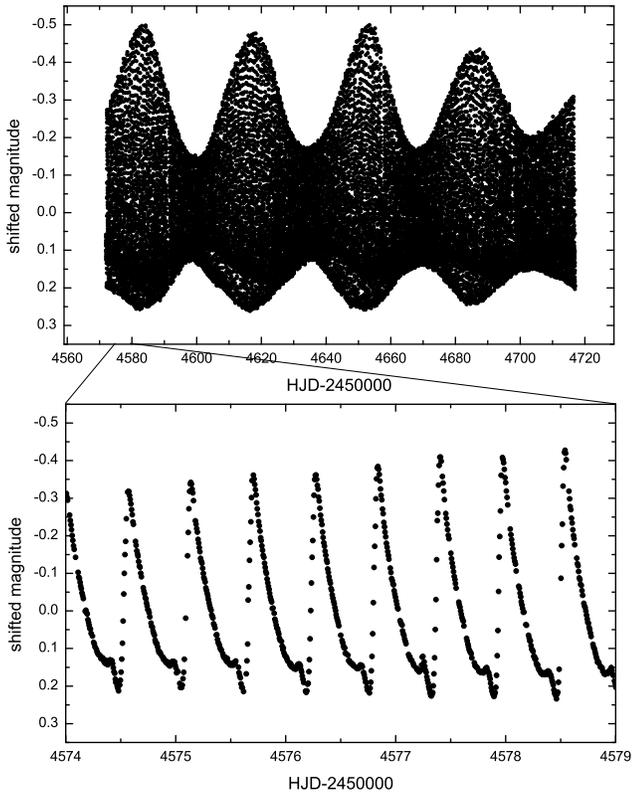}
\caption{The complete data set of CoRoT 105288363 covering more than 4 Blazhko cycles is shown in the top panel. In the bottom panel, a zoom is given to illustrate the continuous sampling of the light curve which is typical for satellite missions like CoRoT.}
\label{lightcurve}
\end{figure}

\begin{figure}
\includegraphics[width=90mm, bb= 10 0 540 350]{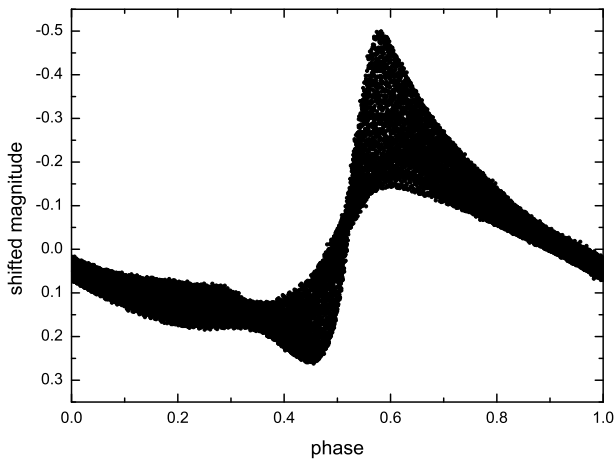}
\caption{Data of CoRoT 105288363 folded with the main period of 0.567d.}
\label{phasedia}
\end{figure}

The CoRoT (COnvection, ROtation and planetary Transits) space telescope is dedicated to stellar seismology and the search for extrasolar planets. It was launched successfully on December 27, 2006 into a polar low-earth orbit with an apogee of 911 and an perigee of 888 km, respectively \citep{auv}. With its unprecedented photometric precision, its continuous time sampling and the possibility to observe a target for up to 150 d it provides an excellent opportunity to study variable stars and especially the long term behavior and stability of their pulsation. \\

As the CoRoT space craft is in a low-earth orbit, the Earth has an influence on the satellite and causes perturbations. The most important of these perturbations are temperature variations due to light/penumbra/shadow transitions and changes in the Earth's albedo, cosmic ray hits during the passage through the South Atlantic Anomaly and satellite attitude variations because of the Earth's gravity field \citep{auv}. These result in instrumental jumps and trends in the data, as well as some strongly deviating points.\\

Due to the above-mentioned effects, the data of CoRoT 105288363 were affected by one significant jump and a slight trend. As the high amplitudes typical for RR Lyrae stars hamper a profound trend correction, and the data sampling would strongly affect the linear fit to the original data, a frequency fit had to be subtracted before the slope of the trend could be determined. The frequency fit contained the main pulsation frequency, its first 10 harmonics and the classical triplet solution around those peaks which describes the mean Blazhko effect of the star. From the residuals we then found the parameters for the jump and trend correction and applied it to the original data. Note that an instrumental trend in the flux data, if not removed, would lead to a spurious change in amplitude due to the conversion to the magnitude scale. This would be fatal for any investigation of the Blazhko effect and its stability, as the amplitudes and their changes are of great interest. Special attention was therefore paid to a correct trend removal, and both jump and trend were removed from the flux data before converting it to magnitudes. Luckily, the trend in the data of CoRoT 105288363 was only marginal and could easily be removed with a linear correction. The fact that the trend was very small in this case rules out the possibility that the observed amplitude changes could originate from contamination with background stars. Figure~\ref{lightcurve} shows the corrected full data set used in this paper, illustrating the full coverage and giving an example of the excellent data quality by zooming into a smaller region. A phase diagram, showing the data folded with the main period, is given in Fig.~\ref{phasedia}. \\

In the framework of the standard CoRoT data reduction pipeline, a 'flag' is attributed to every data point, indicating the quality and reliability of the measurement as well as the nature of the effect that might have affected the measurement in question. As only measurements with flag zero are considered fully reliable, an analysis is usually only performed using those data points. For RR~Lyrae stars, however, it turned out that the rapid brightness changes that occur during rising light caused a lot of measurements to be spuriously classified as non-zero, even though they are high quality measurements. This would result in incomplete coverage and regular gaps in the light curve when only flag=zero points are used. The flag algorithm was therefore not used to exclude measurements, and the light curve was instead cleaned manually from outlying points. The final data set is composed of 20983 useful points, yielding the very high duty cycle of 86 per cent.\\

\section{Analysis and results}
\label{analysis}
\subsection{Fourier analysis}
\label{fourier}
A Fourier analysis of the data was performed using the Period04 software package developed by \citet{len}. It provides simultaneous sine-wave fitting and  least-squares fitting algorithms. The results were thereafter cross-checked with an analysis performed with the iterative sine-wave fitting method \citep{van71} and found to agree within the errors.\\

\begin{figure}
\includegraphics[width=90mm, bb= 10 0 420 560]{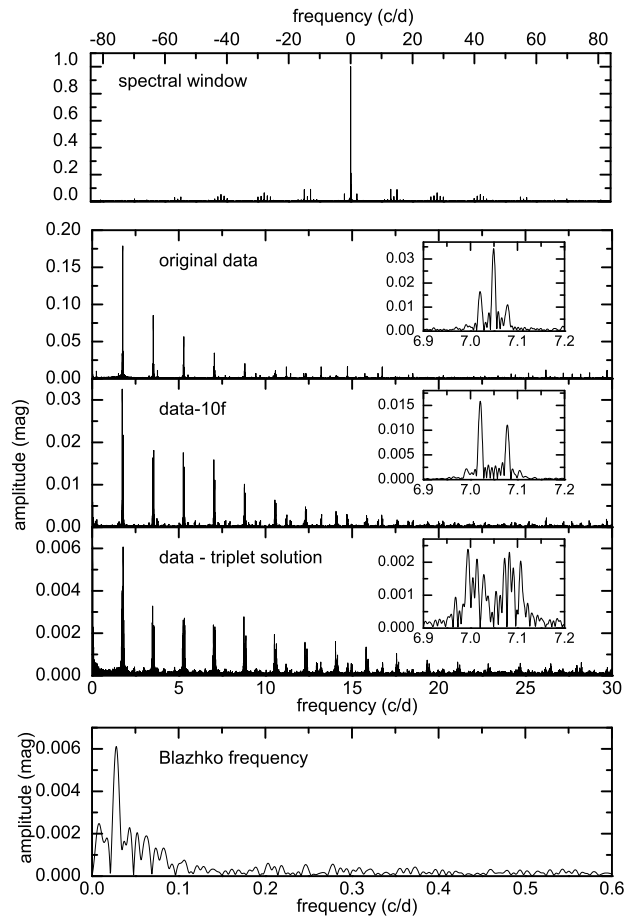}
\caption{Consecutive steps of prewhitening. The top panel shows the spectral window, middle panels give the Fourier spectrum of the original data, the spectrum of the data after prewhitening the main pulsation component and 9 harmonics and the spectrum after subtraction of the triplet solution, respectively. Inserts show an enlargement of the region around $4f_{0}$ to illustrate the triplet structure and the remaining signal. The bottom panel shows the low-frequency region where the Blazhko frequency is directly detected.}
\label{prewhitening}
\end{figure}

The spectral window of the data shows the typical features of high precision satellite data: alias peaks have low amplitudes (smaller than 0.1 normalized amplitude, see upper panel of Fig.~\ref{prewhitening}) and are centered around the orbital frequency and its multiplets. The orbital period of CoRoT is 6184 s \citep{auv}, corresponding to a frequency of $f_{orb} = 13.97$ d$^{-1}$. In addition to the $f_{orb}$ we also find the term $2f_{sid} = 2.0054$ d$^{-1}$, which is caused by the two passages over the South Atlantic Anomaly per day, and its harmonic $4f_{sid}$. The term $f_{sid}=1.0027$ d$^{-1}$ itself is not visible in the spectral window, but its combinations with $f_{orb}$ are present. In summary, we find peaks at $$f_{o,s} = kf_{orb} \pm nf_{sid}$$ with $0 \leq k \leq 6$ and $1 \leq n \leq 7$. The highest peaks are at $(k,n) = (1,\pm1)$, (k,n) = (2,0) and (k,n) = (0,2). \\

The dominant features in the Fourier spectrum of CoRoT 105288363 are the radial fundamental pulsation frequency of 1.7622967 d$^{-1}$ and its numerous harmonics describing the asymmetric light curve shape that is typical for RRab stars. The harmonics were significant up to the 22nd order, when applying a significance criterion of a signal-to-noise ratio of 3.5 for combination frequencies. The noise amplitude calculated in the 2-12 d $^{-1}$ range is 0.07 mmag and is probably still affected by the long-term instrumental drift. Indeed, it is lower (0.04 mmag) in the 50-90 d$^{-1}$ region. The well-known triplet structure with a spacing of the Blazhko frequency (in this case 0.028 d$^{-1}$) that is usually observed in Blazhko stars, is also clearly visible. The triplet could either be caused by a nonradial mode close to the main pulsation mode and its many combinations \citep{bre06}, or it could simply result from the modulation of purely radial pulsation as was recently mathematically described by \citet{szei} and, using a different approach, by \citet{ben09}. The middle panels of Fig.~\ref{prewhitening} show the Fourier spectrum of the original data, after subtraction of $f_0$ and its first 9 harmonics and after subtraction of a fit including the triplet components, respectively. Inserts show a zoom into the vicinity of the 4th harmonic to illustrate the fine structure of the remaining signal.\\

The ephemerides of maximum pulsation light and maximum pulsation amplitude (i.e., Blazhko maximum) found from a classical Fourier analysis of the complete data set are:

\[\rm{HJD}~(T_{\rm{max}})= 2454572.8037 + 0.56744122 \times \rm{E_{pulsation}}\]
\[\rm{HJD}~(T_{\rm{Blazhkomax}})= 2454583.5768 + 35.6 \times \rm{E_{Blazhko}}\]

Note that as the Blazhko effect undergoes strong changes in CoRoT 105288363 (as can already be guessed after a close inspection of the raw data shown in Figure~\ref{lightcurve}). Hence the values above should be understood as mean values of the complete data set. A detailed investigation of the Blazhko period and changes of the Blazhko effect using different approaches will be performed in the subsequent sections.\\

\subsubsection{Multiplet components}
In the past few years it was debated whether only triplet structures occur in Blazhko stars or whether higher-order multiplets are also present, but hidden in the noise of the available data sets. The question was resolved when \citet{hurta} reported the first detection of quintuplets in the RRab star RV UMa using high-precision ground based data obtained during a dedicated campaign. Meanwhile, quintuplets have been found in other RR Lyrae stars as well, for example in the southern star SS~For by \citet{kol09SS}, and thanks to high-precision satellite photometry, even higher-order multiplets are known today, with the record held by a septdecaplet component that has been found in the CoRoT data of V1127 Aql \citep{cha}.\\

In CoRoT 105288363, however, the situation is a bit more complicated. In a Fourier analysis, the strongly changing Blazhko effect and especially its seemingly irregular nature cause close peaks of high amplitude to appear around the dominant peaks. Therefore most multiplet components are hidden, not in the noise, but in an irregular pattern of unresolved peaks that are caused by the cycle-to-cycle changes of the Blazhko modulation. Residual peaks like this were also detected in the two other Blazhko stars observed by CoRoT so far: V1127 Aql \citep{cha} and CoRoT 101128793 \citep{por}. In their cases the amplitudes of the peaks were small, indicating slight changes of the Blazhko cycle. In CoRoT 105288363, however, the amplitudes of the residual peaks are very high, complicating the analysis of the full data set. As a result, the quintuplet peaks become unambiguously detectable only at higher harmonic orders, as their ampitudes do not decrease as rapidly with increasing order as those of the main pulsation components (see also section~\ref{decr}). As it is a valid assumption that if the quintuplets can be detected at higher orders, they are also present at lower orders, also the quintuplet peaks around low orders were included into our fit to the data, even if they lie very close to high-amplitude peaks caused by the varying Blazhko effect. The frequencies included in the final fit to the data of CoRoT 105288363 including the triplet and quintuplet peaks is given in Table~\ref{frequ}. For obtaining this fit, a Fourier sum of the form 
\begin{eqnarray*}
f(t)&=&A_0+\sum_{k=1}^n [A_ksin(2\pi(kf_0t+\varphi_k))\\
&+&A_k^+sin(2\pi((kf_0+f_B)t+\varphi_k^+))\\
&+&A_k^- sin(2\pi((kf_0-f_B)t+\varphi_k^-))\\
&+&A_k^{++} sin(2\pi((kf_0+2f_B)t+\varphi_k^{++}))\\
&+&A_k^{--} sin(2\pi((kf_0-2f_B)t+\varphi_k^{--}))]\\
&+&B_0 sin(2\pi(f_Bt+\varphi_B))\\
\end{eqnarray*}
was used, with $f_B$ being the Blazhko frequency, $A^+$ and $A^-$ denoting the amplitudes of the triplet components, and $A^{++}$ and $A^{--}$ denoting the amplitudes of the quintuplet peaks on the higher and lower side of the main pulsation component, $B_0$ is the amplitude of the Blazhko frequency itself. This is the standard method of fitting the light curves of Blazhko RR Lyrae stars assuming equidistant triplets, extended here to include also equidistant quintuplets and the Blazhko frequency itself, which has only in rare cases been detectable in ground-based data. The Fourier spectrum of the vicinity of the Blazhko frequency is displayed in the bottom panel of Fig.~\ref{prewhitening}.\\

\begin{table*}
 \centering
 \begin{minipage}{140mm}
  \caption{List of frequencies, amplitudes and phases obtained from the analysis of the full data set (all cycles combined), assuming equidistant multiplets. The values should be understood as mean values, as the properties of the pulsation change from cycle to cycle. Errors are 0.00028 mag for the amplitudes and 0.00004 d$^{-1}$, 0.0002 d$^{-1}$ and 0.0014 d$^{-1}$ for the three independent frequencies, i.e., for the main frequency $f_0$, its highest side peak $(f_0-f_B)$ which was used for fixing the spacings, and for $f_1$, respectively. Errors of the phase values obtained from a Monte Carlo simulation with Period04 are given in the $\rm{4}^{th}$ column.}
  \begin{tabular}{@{}lrrrrclrrrr@{}}
  \hline
\multicolumn{2}{c}{f[d$^{-1}$]}	&	A [mag]	&	$\phi$[0:1]	&	$\sigma(\phi)$ MC	& &	\multicolumn{2}{c}{f[d$^{-1}$]}	&	A [mag]	&	$\phi$[0:1]	&	$\sigma(\phi)$ MC	\\
\hline
$f_0$	&	1.76230	&	0.17903	&	0.7418	&	0.0001	&	&	$9f_0+f_B$	&	15.88873	&	0.00167	&	0.9593	&	0.0122	\\
$2f_0$	&	3.52459	&	0.08428	&	0.8827	&	0.0002	&	&	$10f_0-f_B$	&	17.59491	&	0.00141	&	0.3987	&	0.0154	\\
$3f_0$	&	5.28689	&	0.05510	&	0.0622	&	0.0004	&	&	$10f_0+f_B$	&	17.65103	&	0.00111	&	0.1656	&	0.0193	\\
$4f_0$	&	7.04919	&	0.03269	&	0.2444	&	0.0007	&	&	$11f_0-f_B$	&	19.35721	&	0.00093	&	0.5590	&	0.0243	\\
$5f_0$	&	8.81149	&	0.01893	&	0.4277	&	0.0012	&	&	$11f_0+f_B$	&	19.41333	&	0.00085	&	0.3789	&	0.0224	\\
$6f_0$	&	10.57378&	0.00994	&	0.5869	&	0.0022	&	&	$12f_0-f_B$	&	21.11951	&	0.00060	&	0.7324	&	0.0363	\\
$7f_0$	&	12.33608&	0.00594	&	0.7372	&	0.0041	&	&	$12f_0+f_B$	&	21.17562	&	0.00065	&	0.6084	&	0.0338	\\
$8f_0$	&	14.09838&	0.00358	&	0.9055	&	0.0060	&	&	$13f_0-f_B$	&	22.88180	&	0.00026	&	0.9048	&	0.0942	\\
$9f_0$	&	15.86067&	0.00175	&	0.0842	&	0.0120	&	&	$13f_0+f_B$	&	22.93792	&	0.00055	&	0.8338	&	0.0394	\\
$10f_0$	&	17.62297&	0.00061	&	0.2074	&	0.0319	&	&	$14f_0+f_B$	&	24.70022	&	0.00061	&	0.0463	&	0.0307	\\
$11f_0$	&	19.38527&	0.00018	&	0.1304	&	0.1258	&	&	$15f_0+f_B$	&	26.46252	&	0.00052	&	0.2291	&	0.0436	\\
$12f_0$	&	21.14756&	0.00024	&	0.1517	&	0.1032	&	&	$16f_0+f_B$	&	28.22481	&	0.00061	&	0.4271	&	0.0336	\\
$13f_0$	&	22.90986&	0.00036	&	0.2729	&	0.0675	&	&										\\
$14f_0$	&	24.67216&	0.00054	&	0.4239	&	0.0393	&	&		&		&		&		&		\\
$15f_0$	&	26.43446&	0.00052	&	0.5998	&	0.0413	&	&	\multicolumn{5}{c}{significant quintuplet components}									\\
$16f_0$	&	28.19675	&	0.00052	&	0.7459	&	0.0481	&	&		&		&		&		&		\\
$17f_0$	&	29.95905	&	0.00058	&	0.8898	&	0.0387	&	&	$f_0-2f_B$	&	1.70618	&	0.00206	&	0.3443	&	0.0106	\\
$18f_0$	&	31.72135	&	0.00044	&	0.0383	&	0.0578	&	&	$f_0+2f_B$	&	1.81842	&	0.00178	&	0.1541	&	0.0114	\\
$19f_0$	&	33.48364	&	0.00036	&	0.1965	&	0.0656	&	&	$2f_0-2f_B$	&	3.46848	&	0.00256	&	0.5302	&	0.0068	\\
$20f_0$	&	35.24594	&	0.00040	&	0.3853	&	0.0606	&	&	$2f_0+2f_B$	&	3.58071	&	0.00109	&	0.1929	&	0.0175	\\
$21f_0$	&	37.00824	&	0.00031	&	0.5334	&	0.0759	&	&	$3f_0-2f_B$	&	5.23077	&	0.00077	&	0.3088	&	0.0276	\\
$22f_0$	&	38.77054	&	0.00026	&	0.6964	&	0.0843	&	&	$3f_0+2f_B$	&	5.34301	&	0.00231	&	0.4096	&	0.0083	\\
	&		&		&		&		&	&	$4f_0-2f_B$	&	6.99307	&	0.00195	&	0.5242	&	0.0108	\\
\multicolumn{5}{c}{Blazhko frequency}									&	&	$4f_0+2f_B$	&	7.10531	&	0.00191	&	0.6359	&	0.0111	\\
	&		&		&		&		&	&	$5f_0-2f_B$	&	8.75537	&	0.00266	&	0.7795	&	0.0076	\\
$f_B$	&	0.02806	&	0.00614	&	0.4174	&	0.0035	&	&	$5f_0+2f_B$	&	8.86760	&	0.00176	&	0.8317	&	0.0121	\\
	&		&		&		&		&	&	$6f_0-2f_B$	&	10.51766	&	0.00168	&	0.9861	&	0.0119	\\
\multicolumn{5}{c}{significant triplet components}									&	&	$6f_0+2f_B$	&	10.62990&	0.00138	&	0.0387	&	0.0151	\\
	&		&		&		&		&	&	$7f_0-2f_B$	&	12.27996	&	0.00132	&	0.1025	&	0.0178	\\
$f_0-f_B$	&	1.73424	&	0.03296	&	0.9796	&	0.0006	&	&	$7f_0+2f_B$	&	12.39220	&	0.00113	&	0.2433	&	0.0178	\\
$f_0+f_B$	&	1.79036	&	0.02202	&	0.5198	&	0.0010	&	&	$8f_0-2f_B$	&	14.04226	&	0.00139	&	0.2751	&	0.0151	\\
$2f_0-f_B$	&	3.49653	&	0.01613	&	0.0362	&	0.0013	&	&	$8f_0+2f_B$	&	14.15449	&	0.00084	&	0.4335	&	0.0260	\\
$2f_0+f_B$	&	3.55265	&	0.01804	&	0.6578	&	0.0010	&	&	$9f_0-2f_B$	&	15.80456	&	0.00114	&	0.4760	&	0.0185	\\
$3f_0-f_B$	&	5.25883	&	0.01745	&	0.1944	&	0.0013	&	&	$9f_0+2f_B$	&	15.91679	&	0.00064	&	0.6215	&	0.0374	\\
$3f_0+f_B$	&	5.31495	&	0.01391	&	0.8353	&	0.0016	&	&	$10f_0-2f_B$	&	17.56685	&	0.00072	&	0.6594	&	0.0309	\\
$4f_0-f_B$	&	7.02113	&	0.01572	&	0.4035	&	0.0015	&	&	$10f_0+2f_B$	&	17.67909	&	0.00056	&	0.8246	&	0.0355	\\
$4f_0+f_B$	&	7.07725	&	0.01068	&	0.0242	&	0.0022	&	&	$11f_0-2f_B$	&	19.32915	&	0.00051	&	0.8157	&	0.0430	\\
$5f_0-f_B$	&	8.78343	&	0.00979	&	0.5851	&	0.0022	&	&	$11f_0+2f_B$	&	19.44139	&	0.00047	&	0.0232	&	0.0456	\\
$5f_0+f_B$	&	8.83954	&	0.00760	&	0.2074	&	0.0028	&	&	$12f_0-2f_B$	&	21.09145	&	0.00043	&	0.9916	&	0.0524	\\
$6f_0-f_B$	&	10.54572	&	0.00605	&	0.7376	&	0.0034	&	&	$12f_0+2f_B$	&	21.20368	&	0.00039	&	0.2016	&	0.0596	\\
$6f_0+f_B$	&	10.60184	&	0.00513	&	0.3844	&	0.0039	&	&	$13f_0-2f_B$	&	22.85374	&	0.00036	&	0.1764	&	0.0798	\\
$7f_0-f_B$	&	12.30802	&	0.00444	&	0.8836	&	0.0047	&	&	$13f_0+2f_B$	&	22.96598	&	0.00035	&	0.4208	&	0.0629	\\
$7f_0+f_B$	&	12.36414	&	0.00359	&	0.5659	&	0.0053	&	&		&		&		&		&		\\
$8f_0-f_B$	&	14.07032	&	0.00339	&	0.0517	&	0.0066	&	&	\multicolumn{5}{c}{independent mode}									\\
$8f_0+f_B$	&	14.12644	&	0.00249	&	0.7626	&	0.0085	&	&		&		&		&		&		\\
$9f_0-f_B$	&	15.83261	&	0.00232	&	0.2312	&	0.0082	&	&	$f_1$	&	2.984	&	0.0004	&	0.6950&	0.0475	\\

\hline
\label{frequ}
\end{tabular}
\end{minipage}
\end{table*}

\subsubsection{Additional modes}
We also carefully investigated the amplitude spectra to search for additional modes, i.e., frequencies not related with $f_0$ and $f_B$. As in the case of the multiplet components, the strong variation of the light curve shape due to the Blazhko effect is a severe complication in such an analysis. Nevertheless, we found marginal evidence for a mode at about $f_1$=2.984 d$^{-1}$, with an average amplitude of 0.4 mmag. This peak is detected in each Blazhko cycle, with the largest amplitude of 0.8 mmag in the third one. We remind at this point, however, that 17.7 per cent of the observed flux originate from background stars, and that there is a small chance that this signal is caused by the contaminating source.\\

\subsubsection{Modulation parameters}
The properties of the modulation (i.e., multiplet) components, especially the degree of their asymmetry in amplitude is considered a useful tool to constrain the models for the Blazhko effect. The most widely used parameters to describe the appearance of the triplets are the amplitude ratio $R_k=(A_{+})/(A_{-})$ of the two side peaks, and their phase difference $\Delta\phi_k=\phi_+-\phi_-$. These parameters have been calculated for both the triplet and the quintuplet components for each harmonic order and are listed in Table~\ref{parameter}. It is known that about 3/4 of all RRab Blazhko stars in the MACHO database show a triplet with the higher component on the higher frequency side \citep{alcock03}, i.e., $R_k > 1$ . CoRoT 105288363 seems to be part of the rarer class, with the left component having a higher amplitude, and hence $R_k$ values significantly smaller than 1 in most orders. Also in the quintuplets, the left peak is usually higher in CoRoT 105288363. The asymmetry parameter $Q=(A_+-A_-)/(A_++A_-)$, which is similar to $R_k$, was introduced by \citet{alcock03} who find that in their large sample of stars observed in the MACHO project, Q is usually between 0.1 and 0.6, with a peak at 0.3. In our data, the mostly negative values of Q again indicate the dominance of the left side peak. Finally, we include the power difference of the modulation components $\Delta A_k^2=A_{+}^{2}-A_{-}^{2}$. According to \citet{szei} this is the more physically meaningful quantity to measure the asymmetry of the multiplet, as it is connected to the phase difference between the amplitude modulation and the phase modulation, with the triplet being symmetrical if both types of modulation are in phase or if one of the two types occurs in its pure form.\\

\begin{table*}
 \centering
 \begin{minipage}{140mm}
  \caption{Parameters describing the multiplets as explained in the text}
  \begin{tabular}{@{}ccccccccc@{}}
  \hline
\multicolumn{9}{c}{Triplet components}															\\
order k	&	$R_k$	&$\sigma (R_k)$	&$\Delta \phi$	&$\sigma (\Delta \phi)$	&$Q_k$	&$\sigma (Q_k)$	&	$\Delta A^2$	&	$\sigma (\Delta A^2)$	\\
\hline
1	&	0.67	&	0.01	&	-0.460	&	0.002	&	-0.199	&	0.007	&	-6.01E-04	&	1.57E-05	\\
2	&	1.12	&	0.03	&	0.622	&	0.003	&	0.056	&	0.012	&	6.50E-05	&	9.58E-06	\\
3	&	0.80	&	0.02	&	0.641	&	0.004	&	-0.113	&	0.013	&	-1.11E-04	&	8.84E-06	\\
4	&	0.68	&	0.02	&	-0.379	&	0.005	&	-0.191	&	0.015	&	-1.33E-04	&	7.53E-06	\\
5	&	0.78	&	0.04	&	-0.378	&	0.007	&	-0.126	&	0.023	&	-3.81E-05	&	4.91E-06	\\
6	&	0.85	&	0.06	&	-0.353	&	0.010	&	-0.082	&	0.036	&	-1.03E-05	&	3.14E-06	\\
7	&	0.81	&	0.08	&	-0.318	&	0.014	&	-0.106	&	0.050	&	-6.84E-06	&	2.26E-06	\\
8	&	0.73	&	0.10	&	0.711	&	0.021	&	-0.154	&	0.068	&	-5.34E-06	&	1.67E-06	\\
9	&	0.72	&	0.15	&	0.728	&	0.029	&	-0.163	&	0.101	&	-2.58E-06	&	1.13E-06	\\
10	&	0.78	&	0.25	&	-0.233	&	0.049	&	-0.122	&	0.158	&	-7.72E-07	&	7.10E-07	\\
11	&	0.91	&	0.40	&	-0.180	&	0.066	&	-0.050	&	0.223	&	-1.58E-07	&	4.99E-07	\\
12	&	1.08	&	0.68	&	-0.124	&	0.099	&	0.038	&	0.316	&	5.91E-08	&	3.52E-07	\\
13	&	2.09	&	2.47	&	-0.071	&	0.204	&	0.354	&	0.516	&	2.34E-07	&	2.42E-07	\\
\hline
\multicolumn{9}{c}{Quintuplet components}														\\
order k	&	$R_k$	&$\sigma (R_k)$	&$\Delta \phi$	&$\sigma (\Delta \phi)$&	$Q_k$	&$\sigma (Q_k)$	&	$\Delta A^2$	&	$\sigma (\Delta A^2)$	\\
\hline
1	&	0.9	&	0.2	&	-0.19	&	0.03	&	-0.07	&	0.10	&	-1.10E-06	&	1.08E-06	\\
2	&	0.4	&	0.1	&	-0.34	&	0.04	&	-0.40	&	0.12	&	-5.39E-06	&	1.10E-06	\\
3	&	3.0	&	1.1	&	0.10	&	0.06	&	0.50	&	0.14	&	4.75E-06	&	9.65E-07	\\
4	&	1.0	&	0.2	&	0.11	&	0.03	&	-0.01	&	0.10	&	-1.64E-07	&	1.08E-06	\\
5	&	0.7	&	0.1	&	0.05	&	0.03	&	-0.20	&	0.09	&	-3.95E-06	&	1.26E-06	\\
6	&	0.8	&	0.2	&	-0.95	&	0.04	&	-0.10	&	0.13	&	-9.13E-07	&	8.59E-07	\\
7	&	0.9	&	0.3	&	0.14	&	0.05	&	-0.07	&	0.16	&	-4.44E-07	&	6.88E-07	\\
8	&	0.6	&	0.2	&	0.16	&	0.06	&	-0.25	&	0.18	&	-1.23E-06	&	6.43E-07	\\
9	&	0.6	&	0.3	&	0.15	&	0.08	&	-0.28	&	0.23	&	-8.86E-07	&	5.18E-07	\\
10	&	0.8	&	0.5	&	0.17	&	0.09	&	-0.12	&	0.31	&	-2.03E-07	&	3.62E-07	\\
11	&	0.9	&	0.8	&	-0.79	&	0.13	&	-0.04	&	0.41	&	-3.60E-08	&	2.74E-07	\\
12	&	0.9	&	0.9	&	-0.79	&	0.16	&	-0.05	&	0.48	&	-3.07E-08	&	2.31E-07	\\
13	&	1.0	&	1.1	&	0.24	&	0.20	&	-0.01	&	0.56	&	-6.85E-09	&	1.98E-07	\\
\hline

\label{parameter}
\end{tabular}
\end{minipage}
\end{table*}

\subsubsection{Decrease of amplitudes}
\label{decr}

\begin{figure}
\includegraphics[width=90mm, bb= 0 0 560 490]{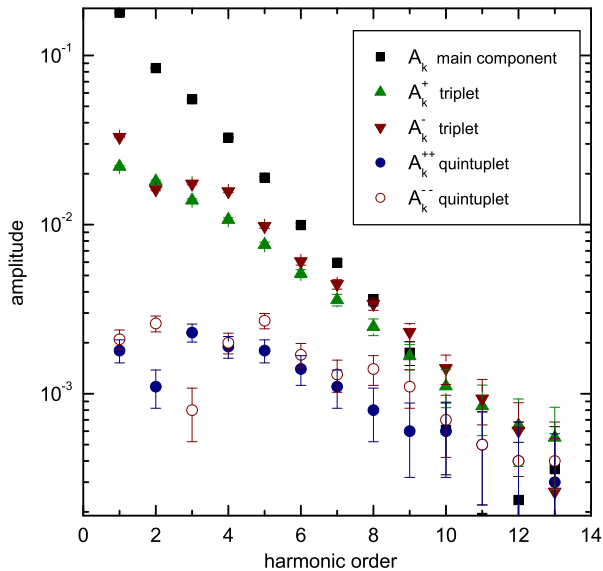}
\caption{Amplitudes of the main pulsation component (squares), the triplet components (triangles) and the quintuplet components (circles) with increasing harmonic order. It is obvious that the amplitudes of the main pulsation components decrease exponentially, while the amplitudes of the triplet and quintuplet components show a less steep, more or less linear decrease. Error bars have been calculated using Monte-Carlo simulations and are smaller than the symbols.}
\label{ampdecr}
\end{figure}

It was first shown by \cite{jur05} that the amplitudes of the harmonics and those of the triplet components behave in different ways with increasing harmonic order. While the main pulsation components (i.e., $f_0$ and its harmonics) decrease exponentially, the triplet components show a much less steep decrease. This was, since then, confirmed for several RR Lyrae stars. In CoRoT 105288363 the result is the same and can be seen in Figure~\ref{ampdecr}. Note that due to the high precision of the satellite data, we are able to show not only the triplets components but additionally also the amplitudes of the quintuplet components in this diagram. This has so far only been done for one star (the prototype RR~Lyrae itself) using the data of the Kepler satellite \citep{kol10a}. Their decrease is more or less linear and even less steep than that of the triplet components. This phenomenon also leads to the effect that at higher harmonic orders, the quintuplet components become the dominant feature after subtracting a fit with the triplet solution, while they are hardly discernible among the disturbing irregular peaks at lower orders. In order 1 the peaks associated with the quintuplet were not visible among the high and numerous other peaks, in orders 2 and 3 only one of the two components was detectable, but with a very low amplitude, from order 4 to 6, both peaks are clearly present but not the highest feature, and from order 7 on, the quintuplet peaks became the dominant signal after subtracting the triplet fit. An example can be seen in the inserts of Fig.~\ref{prewhitening} where the region around $4f_0$ is shown at different steps of prewhitening. After removal of the triplet solution, the 'forest' of peaks caused by the non-repetitive nature of the Blazhko cycles is visible.\\

\subsubsection{Separate analysis of the Blazhko cycles}
As changes in the amplitudes like they occur in CoRoT 105288363 as well as frequency changes are known to cause unresolved peaks to appear in the frequency spectrum of the full data set, a classical Fourier analysis using the complete time string at once is obviously not the best method. We therefore decided to perform a separate analysis of each Blazhko cycle to be able to study the properties of the classical Blazhko multiplets without the presence of dominant disturbing peaks. It was found in this analysis that the amplitudes of the main frequency and its harmonics hardly change from one cycle to the next (they remain within 2 per cent), while the amplitudes of the triplets and quintuplets change by up to 25 per cent. We also observed that, especially in the last Blazhko cycle, the amplitudes of the multiplet features decrease dramatically. The overall picture (i.e., whether the left or the right side  peak is higher), on the other hand, remains constant over all the cycles. Also, higher order multiplet components could be detected in the analysis of the separate cycles while they were hidden under disturbing peaks when analysing the full data set. Depending on the cycle, most or even all septuplet peaks ($nf_0\pm3f_B$) up to order n=18 were detectable as well as some or most nonuplet ($nf_\pm4f_B$) and some undecaplet ($nf_\pm5f_B$) peaks. Triplet and quintuplet peaks emerged clearer at low orders than they did in the full data set, and they were significant up to much higher orders.\\

The frequency analysis of the full data set was therefore only used to find a \textit{mean} pulsation and Blazhko period for the data and to determine the overall properties of the modulation components. \\

\subsubsection{Fourier parameters}
Another approach to avoid the residual peaks caused by the changes in the modulation is to subdivide the light curve into small bins which are hardly affected by the Blazhko modulation and to calculate the amplitudes and phases of $f_0$ and its harmonics for each of them. This method is used to investigate the time-dependent behaviour of the star on a time scale of a few days, reflected by the Fourier parameters.\\

We divided the data set into 73 bins of 2 d duration each (i.e., about 3.5 pulsation cycles), and calculated the Fourier parameters $A_k$ and $\varphi_k$ (with k denoting the harmonic order), as well as their amplitude ratios $R_{k1}=A_k/A_1$ and epoch-independent phase differences $\varphi_{k1}=\varphi_k-k\varphi_1$ in each bin. These Fourier parameters are useful in many aspects. First of all, they describe the shape of the light curve, which for RRab stars is known to change dramatically during the Blazhko cycle. Furthermore, they are a practical tool to compare the properties of synthetic light curves from hydrodynamical models to real data \citep[e.g.,][]{dorfi, smo11}, and \citet{kov95} showed that - for non-modulated stars - it is possible to derive fundamental stellar parameters like the metallicity from the Fourier parameters of RR Lyrae stars. Moreover, \citet{jur02} used the Fourier parameters and their interrelations to test at which Blazhko phase, if at all, the light curves of Blazhko stars resemble those of non-modulated RR Lyrae stars. Also, in the case of a changing Blazhko effect like in CoRoT 105288363, they are useful to compare the different observed cycles in the context of the light curve shape which is given by the parameters.\\

\begin{figure*}
\includegraphics[width=170mm, bb= 0 0 590 800]{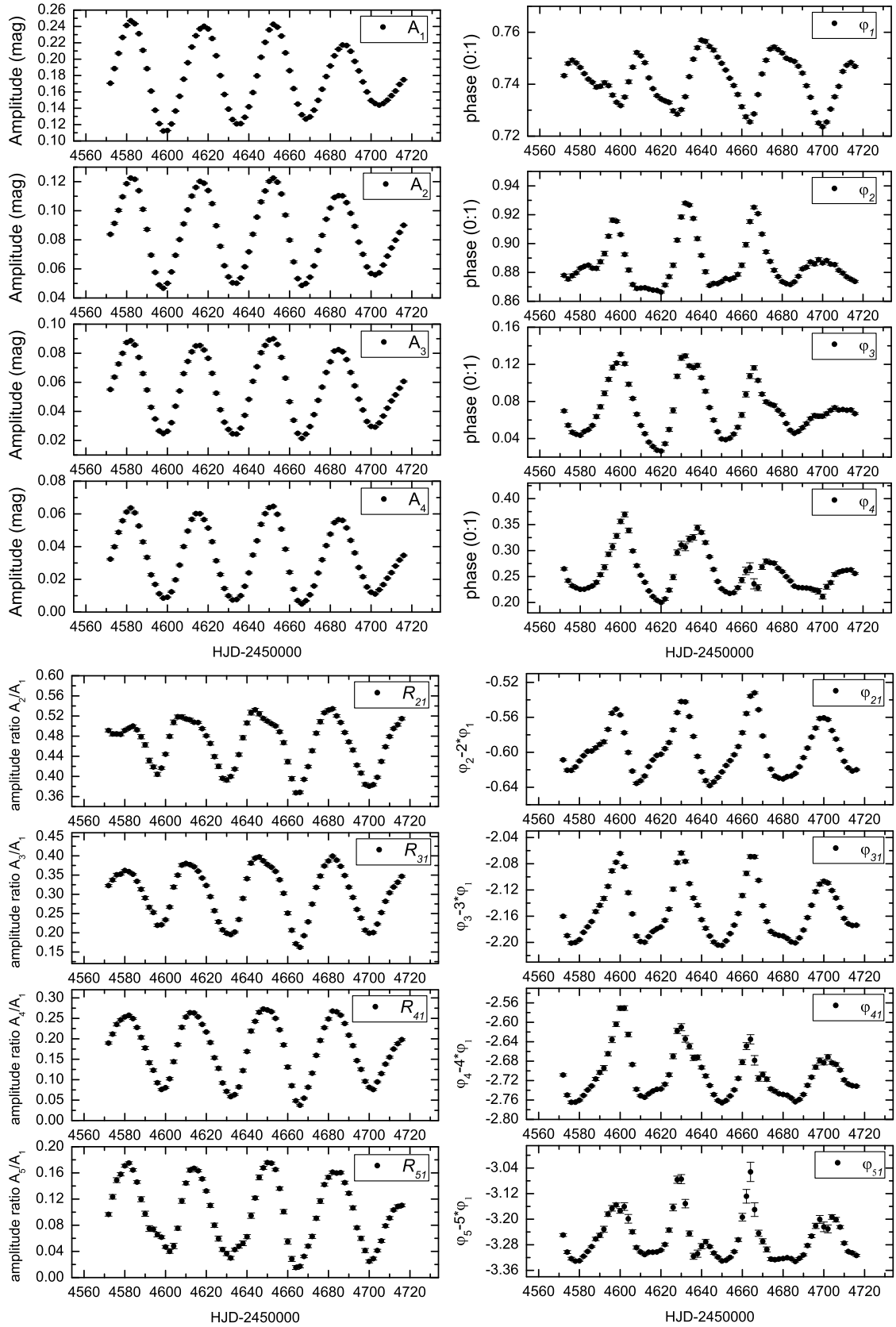}
\caption{Amplitudes, phases, amplitude ratios and phase differences for all 73 bins are shown. Note the error bars barely exceeding the symbol size.}
\label{fp}
\end{figure*}

The Fourier parameters of CoRoT 105288363 are displayed in Figure~\ref{fp}. Some interesting features can immediately be seen: while $A_1$ shows a definite decrease of its variation, indicating a weakening of the Blazhko effect during the duration of the observations, the variation of $\varphi_1$ is getting stronger and sometimes shows double maxima. The other phases also show distinct differences from one cycle to the next, with the fourth observed Blazhko cycle being very weak in all of them. The amplitude ratio $R_{21}$ increases its variation, with flat maxima in the beginning of the observations, and a sharper maximum in the last cycle, and also for the phase differences, none of the Blazhko cycles resembles the other ones exactly.

\subsection{O-C diagrams}
\label{OC}
\begin{figure*}
\includegraphics[width=170mm, bb= 5 0 790 550]{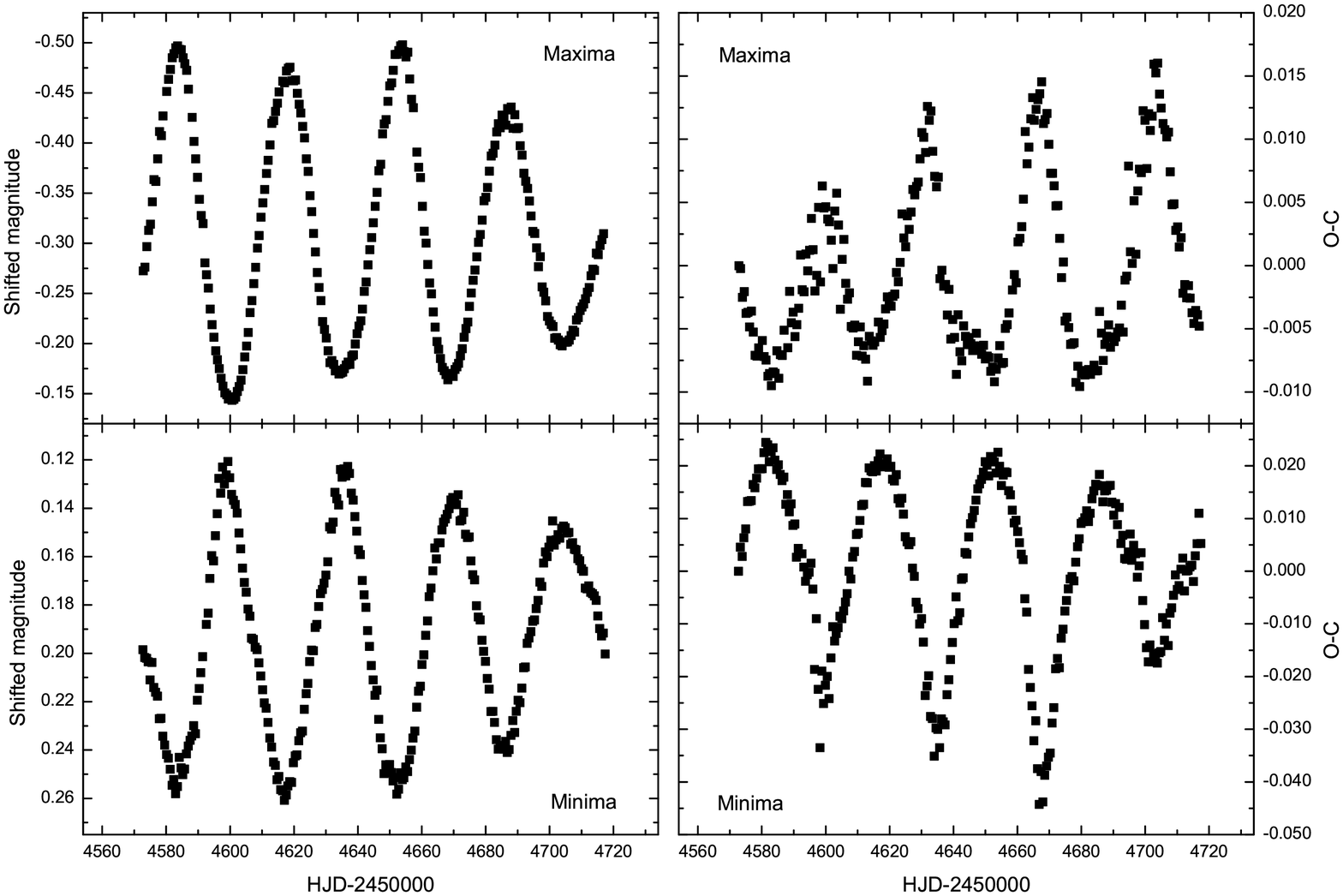}
  \caption{Magnitudes of maximum and minimum light (left panels) and the O-C values of the maxima and minima, respectively (right panels). The strong changes in the Blazhko modulation, both in brightness and O-C variations, are obvious.}
\label{Maxfigure}
\end{figure*}

The O-C analysis is the most classical way to analyse RR Lyrae stars and has nowadays mostly been replaced by Fourier analysis as the preferred method of investigation. In the case of strong changes of the pulsation behaviour, however, when the Fourier spectrum shows a large number of unresolved peaks that hamper the analysis, an O-C diagram might still be a valuable tool to have a closer look at the Blazhko effect. As CoRoT provides continuous time sampling and therefore complete light curves, it is possible to plot an O-C diagram not only for the times of maximum light but also for the minima. As CoRoT 105288363 also shows dramatic changes in the light curve shape, the O-C of minima might hold some additional information. The O-C values of both the maxima and minima are shown in Figure~\ref{Maxfigure}. Additionally, the brightness values of the extrema are shown for comparison.\\

The first obvious observation is that the O-C of maxima and minima show diametrical behaviour, with the minima occurring latest when the maxima are earliest and vice versa. This reflects the fact that CoRoT 105288363 has a skewer light curve during Blazhko maximum than during Blazhko minimum. The next observation is that the O-C of maxima shows an increase in amplitude from one cycle to the next, starting from a value of 0.011 d (i.e., about 16 minutes) peak-to-peak amplitude in the first observed cycle, then increasing to 0.016  and 0.021 and finally to 0.023 d (i.e., 33 minutes) in the fourth cycle, resulting in an increase to about twice the starting value. To obtain these values, five measured light maxima were averaged per Blazhko extremum to balance the scatter. We see a definite strengthening of the Blazhko effect here. Compared to the brightness variations (see left panels of Figure~\ref{Maxfigure} and also the Fourier parameter $A_1$ in Figure~\ref{fp}) this is paradoxical, as the brightness variation indicates a weakening of the Blazhko effect of CoRoT 105288363 with peak-to-peak amplitudes of the maxima (upper left panel of Figure~\ref{Maxfigure}) starting from 0.35 mag and dropping to 0.31, 0.32 and finally to 0.23 mag in the last cycle (with three observed light extrema averaged in each Blazhko extremum to compensate scatter). This diametrical behaviour gives the impression that the energy of the Blazhko modulation might have gradually been converted from amplitude into frequency modulation.\\

The O-C diagram of minimum light offers another intriguing feature: during the first three cycles, the variation gets stronger like for the maxima, from a peak-to-peak amplitude of 0.050 d in the first to 0.051 in the second and 0.062 d third cycle, but the last Blazhko cycle shows only a very weak variation with an amplitude of only 0.035 d, indicating again that also the light curve shape does not repeat exactly. Like it was noted in the Fourier phases $\varphi_3$ and $\varphi_4$ in Figure~\ref{fp}, the fourth cycle seems to deviate strongly from the others. Table~\ref{OCtable} lists the times of maximum light, their magnitude and their O-C values. The full table is available online only.\\

\begin{table*}
 \centering
 \begin{minipage}{140mm}
\caption{Observed maxima and minima of the pulsation of CoRoT 105288363. Columns are: epoch, time of maximum in HJD, relative magnitude, O-C[d], times of minimum, relative magnitude, O-C[d]. The full table is available in the online version only.}
  \begin{tabular}{@{}ccccccc@{}}
\hline
  Epoch    &  Times of maximum & Magnitude  & O-C [d] & Times of Minimum & Magnitude & O-C [d]\\
\hline
0	&	4572.8037	&	-0.2726	&	 0.0000	&	4572.7018	&	0.1985	&	0.0000	\\
1	&	4573.3709	&	-0.2761	&	-0.0002	&	4573.2738	&	0.2017	&	0.0046	\\
2	&	4573.9361	&	-0.2967	&	-0.0025	&	4573.8395	&	0.2020	&	0.0028	\\
3	&	4574.5040	&	-0.3130	&	-0.0021	&	4574.4105	&	0.2035	&	0.0064	\\
4	&	4575.0697	&	-0.3191	&	-0.0038	&	4574.9796	&	0.2110	&	0.0081	\\
5	&	4575.6373	&	-0.3414	&	-0.0037	&	4575.5522	&	0.2038	&	0.0132	\\
6	&	4576.2035	&	-0.3636	&	-0.0049	&	4576.1198	&	0.2141	&	0.0133	\\
7	&	4576.7722	&	-0.3618	&	-0.0036	&	4576.6873	&	0.2161	&	0.0134	\\

... &...&...&...&...&...&...\\
\hline
\label{OCtable}
\end{tabular}
\end{minipage}
\end{table*}

\subsection{Loop diagrams}
\label{loopsect}
\begin{figure*}
\includegraphics[width=170mm, bb= 15 0 740 500]{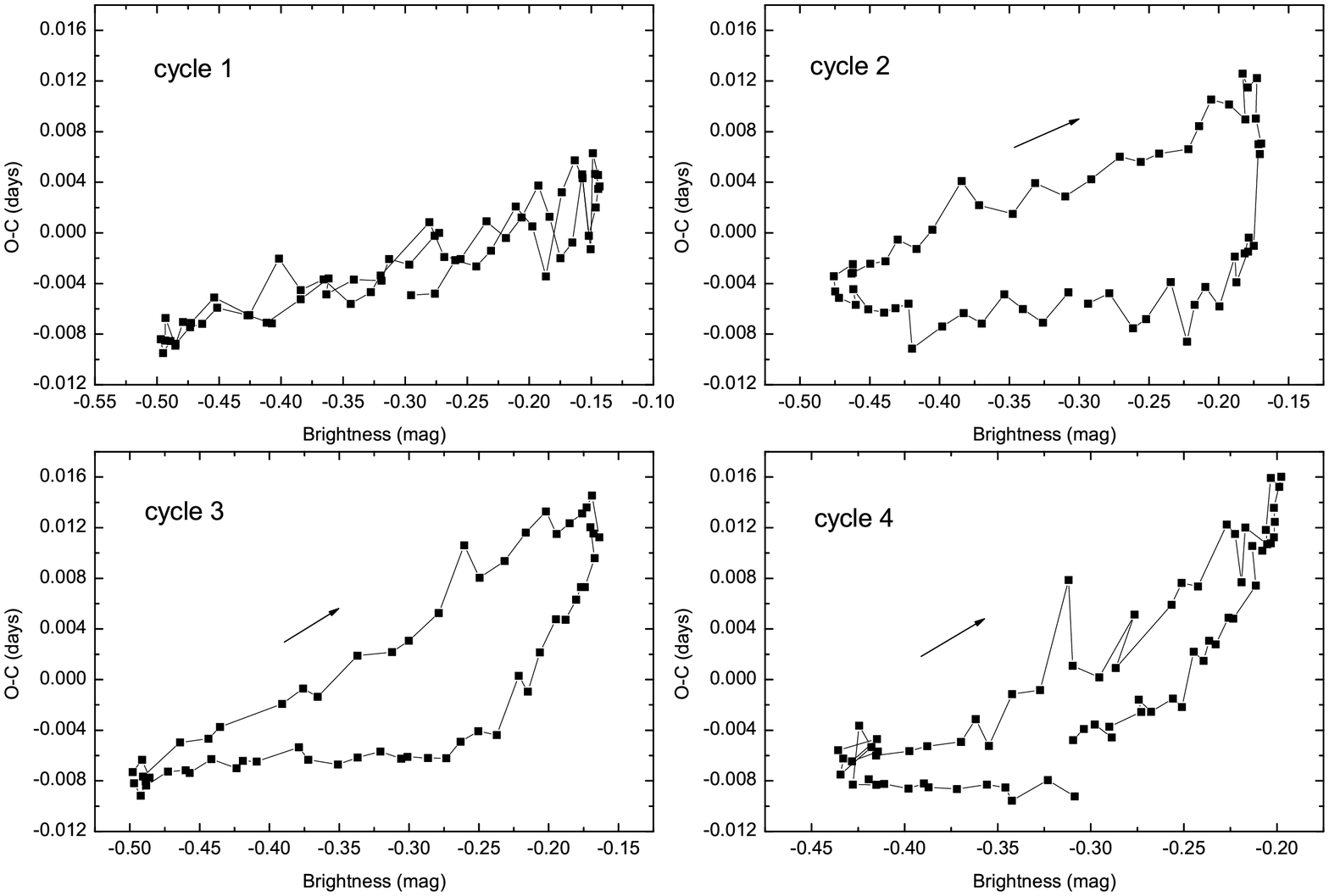}
  \caption{Magnitudes of light maximum versus O-C for the four observed Blazhko cycles}
\label{loop}
\end{figure*}

An important method of diagnostics for the Blazhko effect are the O-C versus brightness, or simply 'loop' diagrams. They represent the contributions of phase and amplitude modulation and the relation between the two. The results for CoRoT 105288363 are given in Figure~\ref{loop} for the four observed cycles separately for better visibility. In the upper left panel, which shows the first cycle, the points lie almost on a single line which indicates that amplitude and frequency modulation are taking place in phase, while the bigger areas that are enclosed by the loops in the second and third cycle indicate that a large phase shift is present. The direction of motion through the diagram is given by the arrows. The fourth cycle appears more like the first one with only a small phase shift. An inspection of the scale of the axes reveals that not only does a phase shift occur and disappear again, but also the strength of the variations changes for both types of modulation, as we already noted in the previous section. A Blazhko cycle was, for the purpose of these plots, defined as from the beginning of the data set to the next time the same Blazhko phase is reached, resulting in 4 observed cycles. When using the definition that a Blazhko cycle starts at Blazhko maximum, the shape of the loops would, of course, be different. However, the fact that the phasing of the amplitude and the phase modulation changes, remains untouched. \\

\subsection{Bump and hump features}
\begin{figure}
\includegraphics[width=90mm, bb= 0 0 550 390]{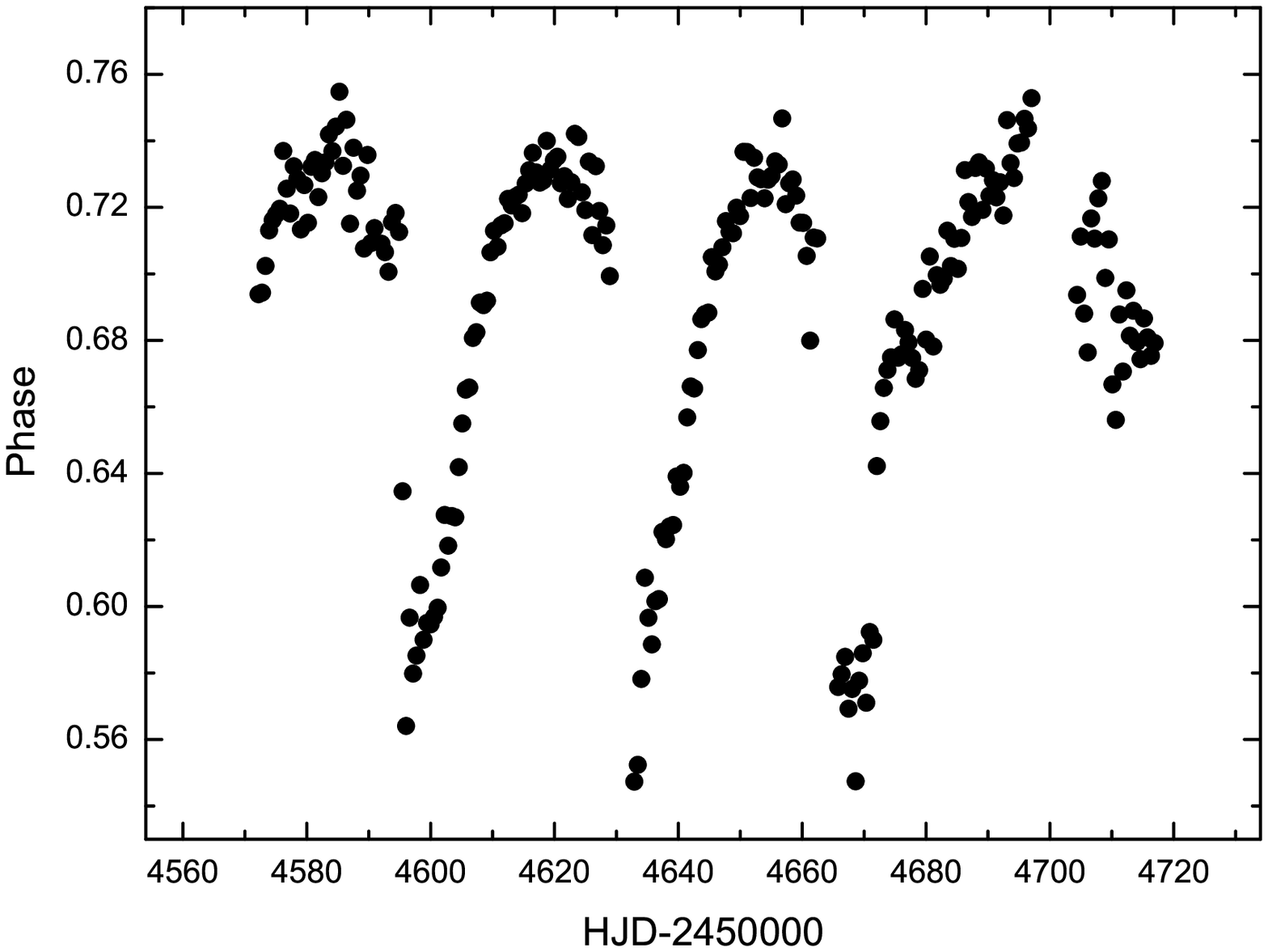}
\caption{Phases of the bump feature}
\label{bump}
\end{figure}

RRab Lyrae stars are known to show a distinct brightness increase -the so-called bump- shortly before minimum light. Unlike in Cepheids, this bump does not show a progression with pulsation period, but it has been noted that it changes its appearance with Blazhko phase in modulated stars. The first detailed investigation on the bump behaviour during the Blazhko cycle has been published by \citep{gug06} for the southern Blazhko star SS~For, using an extended set of ground-based data. With the availability of satellite photometry and therefore uninterrupted observations, it became much easier to see how the bump moves back and forth during the Blazhko cycle. The phases of its maximum for all pulsation cycles observed in CoRoT 105288363 were calculated using the mean period given in section~\ref{fourier} and are plotted against time in Figure~\ref{bump}. The bump occurs later (at a pulsation phase of about 0.72 when zero is set to maximum light) around Blazhko maximum and significantly earlier (at $\varphi$= 0.56) during Blazhko minimum, with a rapid progression back to higher phases which causes a very non-sinusoidal shape of the curve. To exclude the possibility that this effect is just a consequence of the phase modulation, i.e. that it is only caused by the light minima occurring sooner or later depending on the Blazhko phase, the position of the bump was also checked in relation to the adjacent light minimum. The results turned out to be comparable. \\

The other prominent feature that is often observed in RRab stars is the hump or stillstand which occurs during the rising light shortly before maximum light. This feature, which is caused by the so-called main shock, was also investigated quantitatively, although is not very pronounced in CoRoT 105288363 and only hardly discernible at some Blazhko phases. Four consecutive pulsation cycles were combined to obtain a better coverage of the rising branch, and the phase value of the hump was determined by visual inspection. The result follows the same general course of variation as for the bump with later phases at Blazhko maximum and earlier phases around minimum, although, of course, the variation is much smaller. This phase shift of the hump is certainly worth investigating in further studies.\\

\section{Discussions and Conclusions}
\label{disc}
\subsection{Changing Blazhko modulations}
\label{changes}
The fact that the Blazhko modulation in RR Lyrae stars can undergo changes of various types is not totally new. In the literature several reports about phenomena like possible secondary cycles, ceasing and restarting Blazhko modulations, changing values of Blazhko period etc., can be found. The most prominent case certainly is the prototype RR Lyrae with its 4-year cycle which affects the amplitude of the modulation. It was reported that the modulation was barely noticeable in 1963, 1967, 1971 and 1975 \citep{szeidl76}, the years which obviously correspond to the minima of the secondary cycle. But not only those changes in the amplitude of the modulation are known, also variations of the Blazhko period have been
reported. Based on data from the years 2003 and 2004, \citet{kol06} find a Blazhko period for RR Lyrae that is notably shorter than previous values in the literature (38.8 d instead of 40.8 d), while the fundamental period remained roughly the same. It is not known when exactly the change took place.\\

Probably the oldest report about a changing Blazhko effect is the investigation of RW Dra by \citet{balazs57}. The author reports that in 1938, the amplitude of the variation in the O-C diagram caused by the Blazhko effect of RW Dra dropped to about half the value that was measured 3 months earlier (about 1h instead of 2h17m). This reminds us of section~\ref{OC} where we observed the opposite case for CoRoT 105288363: an increase to twice the starting value (33 min instead of 16 min). At the same time, they found that the amplitude modulation of RW Dra got dramatically weaker.\\

Another well-studied example is XZ Cyg. \citet{LaClu04} report that the Blazhko period of XZ Cyg has changed on a time scale of decades, and that these changes are anticorrelated with observed changes of the pulsation period. While in the first half of the 20th century \citep{bla22}, the star was known to have a Blazhko period of 57.4 d, it increased to 58.5
d in the 1960s, while the pulsation period underwent a steep decline in several steps. There was an interval where the Blazhko modulation disappeared for a while, an additional modulation period of 41.6 d was reported to occur at the times when the period of the main modulation was low, and the presence of an overlying long modulation period of almost 10
years was also suspected.\\

The only Blazhko star for which it was possible to unambiguously detect the two modulation frequencies that cause the complex multiperiodic modulation behavior is CZ Lac \citep{sodor09, sodor10}. The two periods have similar amplitudes, so that there is no dominant one, and a strong beating is the result. The periodic decrease in the modulation therefore resembles the cessation of the Blazhko effect in RR Lyrae, and the authors suggest that the cause of the 4-year cycle of RR Lyrae might be the same as in CZ Lac, which would also explain why various different Blazhko periods in the range from 38.8 d to 40.8 d have been reported.\\

Another one of the few stars which have observations available over a sufficiently long time span to investigate the long-term changes of the modulation properties is RV UMa. It was known to show a regular large-amplitude modulation, but based on 90 years of observation \citet{hurta} could find period changes in both the pulsation and modulation period. The range of detected modulation periods was from 89.9 to 90.6 d. An anticorrelation of the changes of the two periods was reported with the relation $dP_{Bl}/dP_{0}=-8.6 \times 10^{4}$. This relation, however, does not seem to very strict, as also an episode of parallel behavior was observed. No change of the modulation amplitude has been reported.\\

An anticorrelation of the changing Blazhko and pulsation period has also been reported for DM Cyg \citep{jur09b}, with a value of $dP_{Bl}/dP_{0}=-1.32 \times 10^4$, based on almost 100 years of observations, while a parallel behaviour of the two periods was found in XZ Dra \citep{jur02} with $dP_{Bl}/dP_{0}=7.7 \times 10^4$. Again, no change of the modulation amplitude has been found. An example of a different type of change is the star RR Gem \citep{sodor07}, with both the amplitude and the period of the modulation undergoing changes. The modulation amplitude showed variations from the undetectable level (less than 0.04mag) to about 0.2 mag, while parallel changes of $P_{Bl}$ and $P_{0}$ were found with $dP_{Bl}/dP_{0}=1.6 \times 10^3$. Both steady and abrupt pulsation period changes were observed, as well as times when the period remained constant.\\

Recently, \citet{jur11} investigated the RR Lyrae variables in the globular cluster M5, and list 6 Blazhko stars for which a change in the modulation period could be determined. Besides this, they also report that in the star V18, the light curve significantly changes from one season to the next while the random period changes if the pulsation period are detected. They state that in this case \textit{the Blazhko effect manifests itself in strong, random, abrupt changes of the light curves, lacking any clear periodic behaviour}. A stochastic and/or chaotic behavior was also suggested for the star MW Lyr \citep{jur08}, a star that, in principle, shows a high degree of regularity in its modulation, but nevertheless reveals significant and non-periodic deviations in the residual light curve. A complex behaviour of the modulation was also suspected by \citet{sodor06} for UZ UMa  because of the unexpected high residuals which still remained even after a fit with two modulation periods was subtracted.\\

As we can see, there are reports about changes in both the amplitude of the Blazhko phenomenon and its period, but the investigations usually suffer from the gaps of ground-based data, larger scatter and from incomplete coverage of the Blazhko cycle. In the homogeneous, uninterrupted and precise data delivered by the CoRoT satellite, we can, for the first time,  perform a detailed study of the changes of the Blazhko modulation. Up to now, in most cases it was unclear when exactly the changes happened or whether they took place abruptly or continuously, while in CoRoT 105288363 we can watch the details of variation taking place.\\

For CoRoT 105288363, an explanation like it was found for CZ Lac (and maybe RR Lyr itself) involving two modulation periods that lead to a beating which strengthes and weakens the cycle can be ruled out, as two modulation periods cannot describe the observed light variation. The variation in the Blazhko effect of  CoRoT 105288363 seem to be of a complex, chaotic nature, like it was proposed for V18 in M5, especially considering the fact that the O-C amplitude is increasing while the brightness modulation decreases (see section \ref{OC}), which is hard to describe with a combination of modulation frequencies. \\

Models which connect the Blazhko effect to rotation, and therefore predict clock-work like behaviour, face significant problems after the detection of the strong cycle-to-cycle variations in the Blazhko effect of CoRoT 105288363. The most promising approaches from the theoretical side are certainly the scenario proposed by \citet{sto, sto10}, as well as the 9:2 resonance with chaos occurring due to the presence of a strange attractor \citep{buchler11}. The 9:2 resonance  can yield the period doubling in some stars or some during specific Blazhko phases, but it seems to be unlikely that it is the sole explanation of the Blazhko effect.\\
On the other hand, the convective cycles proposed in the Stothers models are capable of producing irregular Blazhko cycles. If the light variability of RR Lyr stars is as complicated as the complex behaviour exhibited by CoRoT 105288363 seems to suggest, then both phenomena could coexist. We also note that the Stothers model was critically investigated by \citet{smo11} who concluded that a huge modulation of of the mixing length was required to reproduce the amplitude changes in RR Lyrae.\\

In this context, it is relevant to note the high-amplitude $\delta$ Scuti star CoRoT 101155310 shows small but very clear modulation of the fundamental radial mode \citet{por11}. Though it is still unclear if this is due to star's rotation or an analogue of the Blazhko effect, the thin convective layer  predicted by the physical model of CoRoT 101155310 is an intriguing similarity between  RR Lyr and high-amplitude $\delta$ Sct stars.\\

\subsection{Variations of the bump feature}
Spectroscopically, signatures of emission have been detected during the bump phase \citep{gillet}, strengthening the generally accepted explanation that the bump is caused by a shock wave \citep{hill} that propagates through the atmosphere, resulting from a collision of layers in the deep atmosphere. A change in the mean radius of the star, as it has recently been reported by \citet{jur09} for the Blazhko star MW Lyr could explain different phases of the bump as the run-time of the shock wave through the atmosphere changes and the photosphere is therefore reached earlier or later. A displacement of the region of shock formation during the Blazhko cycle, like it was proposed by \citet{preston}, would also offer an explanation for the changing bump phases.\\

\subsection{Phasing of the modulation}
The dramatic change not only in the strength but also in the appearance of the Blazhko effect certainly provides strong constraints for the models, as an explanation for the Blazhko effect would need to be able to account for strong changes of the phasing of the two types of modulations (amplitude as well as phase modulation) which can be present in a Blazhko star. As we have shown in section~\ref{OC}, the phase variation is getting stronger in CoRoT 105288363, while at the same time the amplitude modulation is getting weaker. Also, we have seen in section~\ref{loopsect} that suddenly a phase shift between the two types of modulation appears and ceases again. Never before have such drastic changes in the Blazhko behaviour of an RR Lyrae star been documented. It is certainly a big challenge for the models to reproduce these phenomena.\\

\subsection{Additional modes}
The excitation of additional modes is an open issue in the modeling of the pulsation of RR Lyr stars. Two frequencies not related with the Blazhko modulation were convincingly found in the case of CoRoT 101128793 \citep{por}. One of them is perhaps related with the period doubling bifurcation, the other supplies the ratio 0.582 with the main pulsation frequency (i.e., 2.119/3.630=0.582). Moreover, the same authors found similar ratio values when rediscussing the cases of V1127 Aql \citep{cha} and MW Lyr (Jurcsik et al. 2008), i.e., 2.8090/4.8254=0.582 and 2.5146/4.2738=0.588, respectively.\\

In the case of CoRoT 105288363 we found a possible additional mode only, i.e., 2.984 d$^{-1}$. The ratio 1.762/2.984=0.591 is very similar to those reported above and this
strengthens our confidence on the reliability of the detection of such small amplitude $f_1$ term. The identification of the additional mode as the second radial overtone, as
in the case of CoRoT 101128793, is still the most plausible explanation.\\

A few similar stars in which frequencies close to the expected value of the overtones appear, were also found in the sample of RR Lyrae stars observed by the \textit{Kepler} satellite \citep{ben10}.

\section*{Acknowledgments}

This research has made use of the Exo-Dat database, operated at LAM-OAMP, Marseille, France, on behalf of the CoRoT/Exoplanet program. KK and EG acknowledge support from the Austrian Fonds zur F\"orderung der wissenschaftlichen Forschung (FWF), project number T359-N16 and P19962-N16. EP acknowledges support from the PRIN-INAF  2010 {\it Asteroseismology: looking inside the stars}. MP, JMB, and RSz acknowledge the support of the ESA PECS projects No.~98022 \& 98114. RSz and JMB are supported by the Hungarian OTKA grant K83790.

\appendix
\section{Comparison with the results of Chadid et al. on CoRoT 105288363}
\label{app}
After finishing the final draft of this manuscript we found out that \citet{cha11} had just published the results of their analysis of the same public data set of CoRoT 105288363. While in some points, our results are in agreement with theirs, there are also some aspects which differ. Those aspects refer to the following items, discussed in a detailed way in the text.

\subsection{Conclusions for models}
\citet{cha11} state that the behaviour of CoRoT 105288363 is a strong support for the scenario described by \citet{sto}. We note, however, that a varying magnetic field modulating the convection is still an unproven hypothesis, and that there is currently no model that can quantitatively describe the interaction between such a field and the turbulent convection.\\ 
\citet{buchler11} use the amplitude equation formalism to study the interaction between the fundamental mode and the 9th overtone in a 9:2 resonance, and find stochastic, chaotic behaviour. The low amplitudes of possible excited overtones and the fact that period doubling is not observed in all Blazhko stars, however, seem to make the 9:2 resonance as sole cause for the Blazhko effect unlikely. On the other hand, the resonance that was found in the simulations of \citet{szabo10} and \citet{buchler11} was startlingly strong, and considering the fact that period doubling is a temporary phenomenon which is present only during certain phases, one might expect that the occurence rate increases when more continuous and accurate satellite data become available.\\
It is, of course, possible that the two effects, the variable turbulent convection as well as the 9:2 resonance, coexist to explain the various observed phenomena. The observed behaviour of CoRoT 105288363 certainly poses a significant problem to all models which require a clock-work like behaviour, and favours explanations that are capable of producing irregular modulation.\\

\subsection{Long-term modulation}
\citet{cha11} report a long-term modulation period of 151 $\pm$ 7 d. We note that the detection of periods exceeding the time base of the availabe data (145 d) has to be taken with great caution. Frequencies which are separated by a value of $\nu_{1}-\nu_{2} < 1/\Delta T $ with $\Delta T$ being the time base of the data, are usually considered unresolved \citep{lou78}. Also, it is well known that artefacts of the Fourier analysis might appear as peaks close to the exact value of $1/\Delta T$. Even though it might be possible under ideal circumstances to predict a long term behaviour from a short data set, we do not think this is the case for CoRoT 105288363. Considering the irregular nature of its modulation which can be seen in both the Fourier parameter variation as well as the O-C variation, it is unlikely that correct prediction of only partially observed long term modulations is possible. The light curve shown in  Fig.~\ref{lightcurve}, top panel, the behaviour of the Fourier parameters  in Fig.~\ref{fp} and the changing shape of the Blazhko cycles in  Fig.~\ref{loop} suggest that a long-term modulation, if periodic, should have a time scale much longer than the time baseline.\\

\subsection{Additional mode}
We find in our analysis an additional, independent mode which was not found by \citet{cha11}. The excitation of additional modes  is a new observational result obtained from the careful  analysis of CoRoT and {\it Kepler} time series. The recurrence  of the same ratio between additional modes and the main pulsational mode is also  discussed in our paper.\\

\label{lastpage}

\end{document}